\documentclass[12pt]{iopart}

\usepackage{epsfig}
\usepackage{amssymb}
\usepackage{latexsym}

\usepackage{color}

\newcommand{\beq}{\begin{equation}}
\newcommand{\eeq}{\end{equation}}
\newcommand{\bea}{\begin{eqnarray}}
\newcommand{\eea}{\end{eqnarray}}

\newcommand {\Kalpha} {{\cal  K}^{(s)}(\alpha)}
\newcommand {\Kcalpha} {{\cal  K}_{(c)}^{(s)}(\alpha)}

\begin{document}

\title[]{Large random correlations in individual mean field spin glass
  samples}

\author{Alain Billoire$^1$, Imre Kondor$^2$, Jovanka Lukic$^3$, 
Enzo Marinari$^3$}

\address{ $^1$ Institut de Physique Th\'{e}orique, CEA Saclay and
  CNRS, 91191 Gif-sur-Yvette, France\\ $^2$ Department of Physics of
  Complex Systems, E\"otv\"os University, Budapest, Hungary\\ 
  $^3$ Dipartimento di Fisica, IPCF-CNR and INFN,
  Universit\`a di Roma ``La Sapienza'', P. A. Moro 2, 00185 Roma,
  Italy}
\ead{alain.billoire@cea.fr\\ kondor.imre@gmail.com\\ jovanka.lukic@gmail.com
  \\ enzo.marinari@uniroma1.it}
\begin{abstract}
  We argue that complex systems must possess long range correlations
  and illustrate this idea on the example of the mean field spin glass
  model. Defined on the complete graph, this model has no genuine
  concept of distance, but the long range character of correlations is
  translated into a broad distribution of the spin-spin correlation
  coefficients for almost all realizations of the random
  couplings. When we sample the whole phase space we find that this
  distribution is so broad indeed that at low temperatures it
  essentially becomes uniform, with all possible correlation values
  appearing with the same probability. The distribution of
  correlations inside a single phase space valley is also studied and
  found to be much narrower.
\end{abstract}


\section{Introduction\label{S:INT}}

Spin glasses~\cite{MEPAVI} and their (in many respects well
understood) mean field versions, like the Sherrington-Kirkpatrick (SK)
model~\cite{SK}, have emerged in the last decades as an interesting
paradigm for complexity: indeed, they represent one of the few
approaches to complexity where one can gain a good analytical and
mathematical control. In these systems, with competing quenched random
couplings inducing disorder and frustration, one has a phase
transition to a low temperature spin glass phase. The critical
temperature $T_c$ is characterized by a divergent correlation length,
but, more remarkably, the whole ordered phase in the range $0\leq
T\leq T_c$ is critical ($T$ is the temperature of the system): the
presence of an infinite number of equilibrium states implies that, at
all temperatures in the spin glass phase, the state to state
transitions generate an infinite correlation length.

Broken continuous symmetries also lead to the appearance of zero modes
all through the ordered phase of ordinary translationally invariant
systems. No such continuous symmetry is present in the Ising spin
glass however, and the mechanism of the generation of long range
correlations is completely different from, say, the one induced by the
long wavelength spin waves in the Heisenberg model. We believe that
the multiattractor structure of the spin glass phase and the long
range correlations generated by it are characteristic of many complex
systems, and it is in this context that we wish to place the present
work.

The idea that the spin glass phase is, in some sense, ``soft" has been
around from the very beginning of the field. The stability analysis of
the Parisi solution for the SK model provided a proof for the
existence of zero modes in the replica symmetry broken phase
~\cite{DDK}. The authors went on to calculate the various propagators
in the first (Gaussian) correction to the mean field approximation of
replica field theory, and found different power law like asymptotic
behaviors, depending on the value of the replica
overlap~\cite{DDK84,DDK85}. Somewhat later, they presented a formally
exact theorem for the existence of zero modes all through the replica symmetry
broken phase~\cite{KDD86}, while further work clarified the connection
between the spin glass propagators and the various overlap
correlations that can be defined in the model, and worked out the long
wavelength asymptotics of these correlation functions in the Gaussian
approximation~\cite{DDKOTE,DDKOTE98}. Extensive numerical calculations
found convincing evidence for these long range correlation functions
in low dimensional spin glass models, although they also demonstrated
the strong renormalization of some of the correlation exponents, with
respect to the Gaussian approximation~\cite{Appropriate}. All the
analytic studies cited so far took the ``many valleys", or ``replica
symmetry breaking" picture as their starting point. Remarkably, soft
excitations and power law like behavior for the correlation functions
are also an integral part of the rival droplet theory~\cite{FHBM}. It
seems therefore that, at least as far as the algebraic character of
the long distance behavior of correlation functions is concerned,
there is a rare general agreement between the various schools of
thought.

Most of the analytic as well as numerical studies of spin glass
correlations have focused on the behavior of quantities averaged over
the quenched disorder~\footnote{Notable exceptions are the series of
  works~\cite{SICAMA2000,SCCM2001,HUKIBA2002,COMAMA2002} investigating
  the eigenvalue spectrum of correlations in the individual samples as
  a tool to uncover the character of the ordered phase.}. In the
present work we propose to consider the simplest spin-spin
correlations without averaging over the randomness. While the average
of this object must be trivial, the individual samples must display a
highly nontrivial correlation structure that has so far been dismissed
as a legitimate object of study, on account of its random, chaotic
character.  We wish to study this object in the SK model not for its
own sake, however, but as perhaps the simplest manifestation of the
appearance of strong correlations generated by a complex phase space
structure.  Looking at the model from the point of view of real world
complex systems whose dynamics depends on competing interactions
(neural networks, complex optimization, the economy, the financial
network, the power grid, traffic, etc.) we have to realize that these
are usually large, but not infinitely large This means that their
various subsystems cannot necessarily be expected to realize all the
different samples of the random couplings, which invalidates the
rationale for averaging over the randomness. In this context, it is
therefore a most relevant question whether the various spin glass
samples display some common characteristics that may be regarded as
typical, and may be expected to show up also in other complex
systems. In particular, we wish to look into the distribution of
correlations and see how general the appearance of large random values
is.

Let us start by recalling some facts about criticality and
correlations.  Ordinary, non-complex systems display short range
correlations. Beyond a certain length scale the various parts of the
system become independent of each other. This is the property behind
ordinary thermodynamic behavior (e.g. extensive internal energy,
thermal fluctuations of extensive quantities proportional to
$\sqrt{N}$, etc.). An equivalent statement is that the probability
density functions for the whole system approximately factorize
into distributions belonging to (sufficiently large) subsystems. In
the renormalization group (RG) framework, one has a renormalization
group flow to a Gaussian, or high temperature fixed point.  When,
however, a system is complex and it ``is more than the sum of its
parts'', the above properties do not hold: in particular,
correlations must be long ranged (see also the recent discussion
in~\cite{KOZJ}).

An example of this type of behavior is the critical
state. Correlations in the critical state decay like a power of the
distance. In a usual, translationally invariant critical system, they
decay monotonically. As for the ordered phase of an ordinary,
translationally invariant system, its correlations, calculated over
the full Gibbs ensemble, decay to a constant, and it is only within a
pure phase that clustering holds and connected correlations vanish for
large distances.  We have said that the ordered phase of spin glasses
is in some sense critical: it is reasonable then to ask what the
behavior of correlations is in this case. What correlations do after
averaging over randomness is well understood at the level of the
Gaussian approximation~\cite{DDKOTE98}.  Averaged squared correlations
decay as power laws, and they decay monotonically. They show this
behavior both in the full ensemble, and in the pure states, the
individual \textit{valleys}.  This implies, and we are getting here
close to the point we want to discuss, some kind of long-range
behavior also for the correlations in a given sample: the average
value of the self-overlap of correlations inside a given valley, taken
at a fixed long distance, cannot be large without at least some of the
sample correlations, taken at the same distance, being large. Hence,
the correlations in the individual samples must be long ranged, in
some sense. In what sense? The system is not translationally
invariant, therefore there is no reason to expect that the
correlations in the individual samples behave monotonically. They may
drop over small distances, becoming large again somewhat farther away,
may change sign a number of times along the way, and start to decay
(in absolute value) for even longer distances. If these correlations
are to be long ranged, they must take up large values with some
probability even for very long distances.

This picture of the behavior of correlations in the pure phase of a
spin glass says that large values of these correlations tend to
cluster at small distances, but there is no monotonicity, and
occasionally there are large correlations even between very far away
partners. The probability of finding a strongly correlated pair at
large distances is decaying with the distance, but there will always
be strongly correlated pairs separated by very long
distances. Averaging over these correlations one should be able to
recover the known results (bearing in mind that the
results~\cite{DDKOTE98} were derived in the Gaussian approximation).
The existence of these long ranged correlations in every sample
explains the curious behavior in the thermodynamic limit: adding
another layer to even a very large system may reorganize the
equilibrium distribution of spins inside the bulk, because large
correlations may be generated between the old spins and the ones in
the newly added layer. The existence of large correlations is then
seen to be intimately related to the so called ``chaos in system
size'' ~\cite{NS}.  For the same reason, due care must be taken when
specifying boundary conditions in a simulation for spin glasses, and
also when defining a real space renormalization procedure, for
example. A broad range of correlations at all length scales has been
found to appear in hierarchical lattices~\cite{ARABER09}.

In this note we would like to understand what all this may imply for
the Sherrington-Kirkpatrick (SK) model.  We regard this model as the
simplest laboratory where the phenomenon of large, random
correlations, that we conjecture to be a generic feature of complex
systems, can be studied.

Let us first consider a simple ferromagnetic Ising model (in a finite
number of dimensions). Here if we are above the critical point the
correlations decay exponentially. The distribution of correlations in
such a situation is extremely sharply concentrated around zero: the
number of correlations that are essentially different from zero is of
order $N\;z$, where $N$ is the number of spins in the system and $z$
is roughly the number of spins within the correlation radius, whereas
the total number of correlations is $N(N-1)/2$. So most of the
correlation values are small, and their distribution has a sharp peak,
with the total weight outside the peak being of order $1/N$.  As for
the corresponding long range (mean field) model, correlations here are
basically trivial.  The correlation between two distinct spins is
$1/(N(T-1))$.  If we plot the histogram of all these correlations we
get ${\cal O}(N)$ points at the value one, and ${\cal O}(N^2)$ points
at the value $1/(N(T-1))$.
 
What do we expect exactly at the critical point? In the corresponding
short range model correlations become long ranged at this point, but
they still go to zero for large distances. Nevertheless, because of
the long range nature, the number of large correlations will now be
larger, and the central peak will be broader. In the infinite range
model the correlation between two distinct spins is now $1/{\cal
  O}(\sqrt{N})$. As we go below the critical point, the Gibbs ensemble
splits into two pure states. If we calculate the correlations over the
whole ensemble, they will not go to zero for large distances. If we
calculate the connected correlations inside one pure state, however,
we will again find that they drop to zero at large distance, and in
the short range model they decay to zero exponentially again.

Here, we will try to understand how this discussion applies to the SK
model, and eventually find that the situation there is to a large
extent analogous to the case of the ferromagnet.  We will use
numerical simulations to measure, for a given quenched random sample
$s$, the spin-spin correlation function ${\cal K}^{(s)}_{ij}\equiv
\langle \sigma_i\sigma_j\rangle^{(s)}$ and its connected part in
different situations (always for the Sherrington-Kirkpatrick model),
and we will discuss the physical picture that emerges from these
measurements. In section \ref{S:MET} we define the model, describe the
methods we use and formulate the questions we want to answer.  In
section~\ref{SS:ALL} we discuss correlations measured when visiting
all of the phase space. In section~\ref{SS:ONE} we discuss
correlations in a single free energy valley, and in
section~\ref{SS:CON} we analyze their connected parts. In
section~\ref{S:CON} we draw our conclusions.

\section{Models, methods and questions\label{S:MET}}

In the following we will consider the Sherrington-Kirkpatrick
model~\cite{SK}, i.e. the mean field model (on a fully connected graph)
for Ising spin glasses. The Hamiltonian has the form
\begin{equation}
\label{eq:sk}
{\cal H} \equiv - \frac{1}{\sqrt{N}} \sum_{1\leq i<j \leq N} 
\sigma_i\; J_{i,j}\; \sigma_j\;,
\end{equation}
where the $\sigma_i$ are Ising spins that can take the values $\pm 1$
and are labeled by an index $i$ running from $1$ to $N$. The
couplings $J_{i,j}$ are quenched independent identically distributed
random variables that define the strength of the interaction between
couples of spins. We will assume that their disorder expectation value
$\overline{J_{i,j}}$ is equal to zero, and that
$\overline{J^2_{i,j}}=1$: under these conditions the system undergoes a
phase transition to a spin glass phase~\cite{SK,MEPAVI} at $T_c=1$.
In all our numerical work the couplings will take the values $\pm 1$ with
probability one half. We will always work with setting the magnetic field
equal to zero.

As we have discussed in the introduction, we will analyze numerically
the spin-spin correlation functions in the system. After averaging
over the quenched couplings, the gauge invariance of the system makes
them trivial~\cite{TOULOU1977,MAPARI}, but looking at
individual disorder samples one can still read out some interesting
information. We use Monte Carlo numerical simulations (about which we
will give more details in the following) to compute the spin-spin
equal time correlation functions (thermal averaged at temperature $T$)
in the disorder realization (sample) labeled by the index $s$
\begin{equation}
\label{eq:k}
{\cal K}_{ij}^{(s)} \equiv 
\langle \sigma_i \sigma_j\rangle^{(s)}\;,
\end{equation}
and their connected parts
\begin{equation}
\label{eq:kc}
{\cal K}_{(c)\;ij}^{(s)} \equiv 
\langle \sigma_i \sigma_j\rangle^{(s)}_c
 \equiv 
\langle \sigma_i \sigma_j\rangle^{(s)} -
\langle \sigma_i\rangle^{(s)} \langle \sigma_j\rangle^{(s)}\;.
\end{equation}
The behavior of these functions is a priory non trivial, and can
give an interesting insight into the physics of the system. 

Normally two point correlation functions are analyzed as functions of
the distance between the two spins: correlations decay (or go to a
constant value) at large distance, and the rate (or the absence) of
this decay is of paramount importance for understanding the behavior
of the system. In our mean field, Sherrington-Kirkpatrick
case~ \footnote{When defining a mean field model on a random diluted
  graph, a distance can be introduced, although its interpretation as
  a bona fide physical distance is not at all straightforward.}, there
is no notion of a spatial distance between the different spins: all
spins directly interact with all other spins, making the effective
dimensionality of the model infinite. Since we cannot use a bona fide
distance to detect a decay of correlation functions, we take a
different approach: we sort the correlations in order of decreasing
magnitude (we do that in each single sample of the random quenched
noise, otherwise, as mentioned before, we would get a trivial answer)
and we study the functional form of this decay (in order to better
connect with the integrated distribution function, we in fact organize
the correlations in order of increasing magnitudes, but obviously both
procedures are equivalent).  In other terms, we are considering the
$N(N-1)/2$ measured numbers ${\cal K}^{(s)}_{ij}$ as the empirical
distribution function of correlations in a given sample.  Our goal is
to understand how many spins are ``really'' interacting with each
other: is this interaction decaying ``fast'' (i.e. does a typical spin
interact with a small number of other spins or does it correlate with
a large group of spins)? This concept is, in some sense, the best
proxy for a real distance.

We introduce the probability $\pi(X)$ that the correlation
$\langle \sigma_i \sigma_j\rangle$ is equal to $X$ (all these
quantities are defined for a single disorder sample, but we omit the
index $s$ when it is not strictly necessary), and the integrated
distribution function
\begin{equation}
\label{eq:int}
\phi\left(X\right) \equiv \int_{-1}^X dY\;\pi(Y)\;.
\end{equation}
$\pi(X)$ depends on the temperature $T$.  Given a set of equilibrium
measurements, one can evaluate the empirical estimator for a generic
function $g$ of the correlation functions:
\begin{equation}
\label{eq:estimator}
\frac1{N(N-1)}\sum_{i\ne j}
g\left(\left\langle
\sigma_i \sigma_j
\right\rangle\right)
\simeq
\int_{-1}^1 dY\;\pi(Y)\;g(Y)\; .
\end{equation}
For example
\begin{eqnarray}
\label{eq:estimatorx2}
\int_{-1}^1 dY\;\pi(Y)\;Y^2
&\simeq&
\frac1{N(N-1)}\sum_{i\ne j}
\left\langle
\sigma_i \sigma_j
\right\rangle^2
=
\frac1{N(N-1)}\sum_{i\ne j}
\left\langle
q_i q_j
\right\rangle\\
\nonumber
&\simeq &\left\langle \Bigl( \frac{1}{N}\sum_{i}
q_i \Bigr)^2
\right\rangle \equiv <q^2>
\; ,
\end{eqnarray}
where we have defined the overlap at site $i$ of two clones (also
known as ``real replicas'') of the system, $\sigma_i$ and $\tau_i$
(two independent copies of the system whose Hamiltonians contain the
same realization of the quenched disorder) as $q_i\equiv \sigma_i
\tau_i$. In the infinite volume, thermodynamic limit, in the high
temperature, paramagnetic phase the distribution $\pi(X)$ will be a
trivial Gaussian centered around zero, while it will be non trivial in
the low temperature, spin glass regime. Note that a broad distribution
of the correlations implies a broad distribution of the overlaps.

We analyze this system by numerical simulations. The complex structure
of a mean field spin glass, and the presence of free energy valleys
whose number diverges with $N$, separated by free energy barriers
whose height diverge with $N$, make analyzing the phase space a
difficult task. The use of an optimized dynamics is mandatory, and we
use the parallel tempering algorithm~\cite{PT,PT1}, that is one of the
most effective updating methods available.  We have used systems of
sizes ranging from $N=128$ up to $N=2048$, and investigated
temperatures in the range from $T=0.4$ up to $T=1.3$.

We analyze two different kinds of situations. In both cases, we start
from very well equilibrated spin configurations obtained with a very
long ($1.4 \ \times 10^6$ Monte Carlo sweeps of the system) parallel
tempering simulation.  In the first case (the ``all states'' case), we
have fairly sampled all the phase space: this means that we have
inspected all the ``states-to-be'' that manifest themselves in finite
volume: the finite energy valleys separated by very high, but (for
finite $N$) finite, barriers. This is achieved using the parallel
tempering algorithm again. A second part of our analysis (the ``one
state'' case) is based on a dynamics that only inspects a single
valley: this allows us to determine the features of the decay of
correlations inside a single state (to be). A lazy dynamics that only
inspects a single valley is easy to build: one only needs a large
enough value of $N$, and a local, non optimized dynamics (we take the
usual local Metropolis algorithm).  One starts from a well thermalized
configuration and runs the local dynamical updating scheme for a limited
amount of time (in order to keep the probability of a transition to a
different valley negligible).  Monitoring the probability density
of the overlap, $P(q)$, we are able to check that we are really not
leaving the valley from which we started. An a priori check can be also obtained by considering
 the measured distribution of relaxation time $\tau_J$   (the time it takes for the system
to forget it starting point) for the Metropolis dynamics, as a function of $T$ and $N$. If the system is large enough  $\tau_J$
 is extremely
long, with~\cite{Young} $\overline{\ln \tau_J} \propto N^{1/3}$.

In all these situations we measure the $\frac12 N(N-1)$ correlations
${\cal K}_{ij}^{(s)}$ and the $N$ magnetizations, i.e. the expectation
values of the spin in the sample $(s)$:
\begin{equation}
m_{i}^{(s)}\equiv \left\langle \sigma_i \right\rangle^{(s)}\;.
\end{equation}
In the ``all states'' simulation, the sampling is very good and in
particular we sample a given state and the state obtained by a global
spin reversal with the same frequency. Accordingly we have
$\left\langle \sigma_i \right\rangle^{(s)}$ compatible to zero also in the
spin glass phase. (The non zero $m_{i}^{(s)}$ used in analytical
works are defined in the thermodynamic limit as the zero magnetic
field limit $\lim_{H\to 0}\lim_{N\to\infty} \left\langle \sigma_i
\right\rangle^{(s)}$.)

We also measure the connected correlation functions ${\cal
  K}_{(c)\;ij}^{(s)}$.  These quantities are always defined as thermal
averages for a given sample of the quenched disordered couplings.  We
then sort the $N(N-1)/2$ numbers ${\cal K}_{ij}^{(s)}$ and ${\cal
  K}_{(c)\;ij}^{(s)}$ in increasing order of magnitudes. We define a
normalized rank $\alpha$ $\in[0,1]$ dividing the rank by $N(N-1)/2$,
and call $\Kalpha$ and $\Kcalpha$ the values of the ordered
correlations as a function of the normalized rank.

\section{Results\label{S:RES}}


\subsection{The correlation functions for the full phase space\label{SS:ALL}}

\begin{figure}
\centering \includegraphics[width=0.6\textwidth,angle=270]{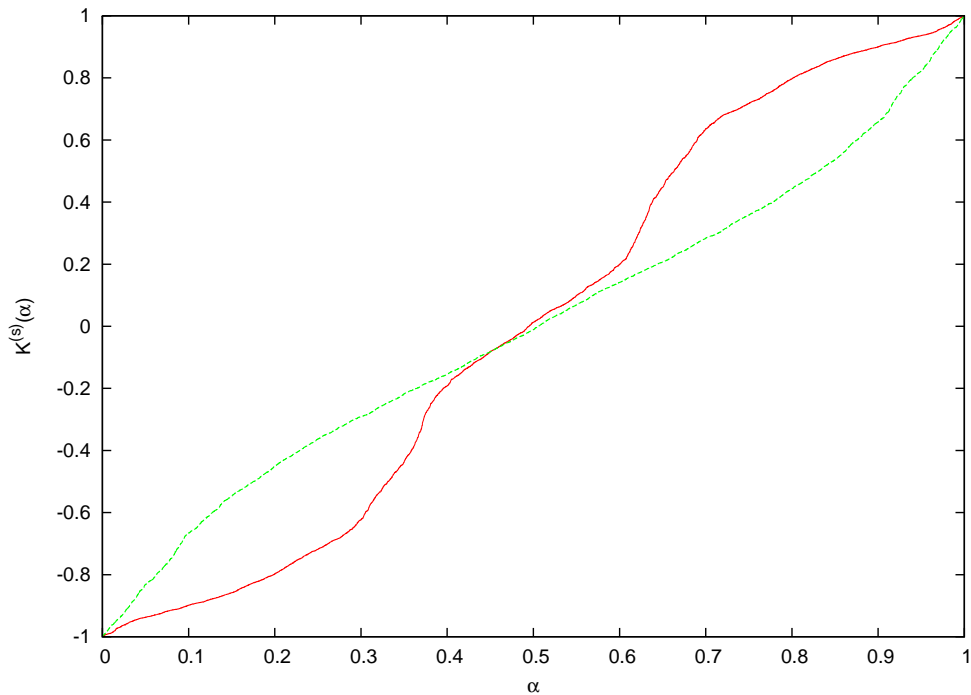}
\caption{${\cal K}^{(s)}(\alpha)$ as a function of the normalized rank
  $\alpha$.  Here $\Kalpha$ has been computed by averaging over all
  phase space, with $N=128$ and $T=0.4$. The two curves are for two
  distinct disorder samples.\label{figureK128}}
\end{figure}

We start by discussing our results for the correlation functions
${\cal K}_{ij}^{(s)}$: here there has been no subtraction of the
magnetizations. We show in Fig.~\ref{figureK128} $\Kalpha$ as a
function of $\alpha$ for two different samples of the quenched
disorder. Here we are working with relatively few sites and at low
temperature: $N=128$ and $T=0.4$.

\begin{figure}
\centering \includegraphics[width=0.6\textwidth,angle=270]{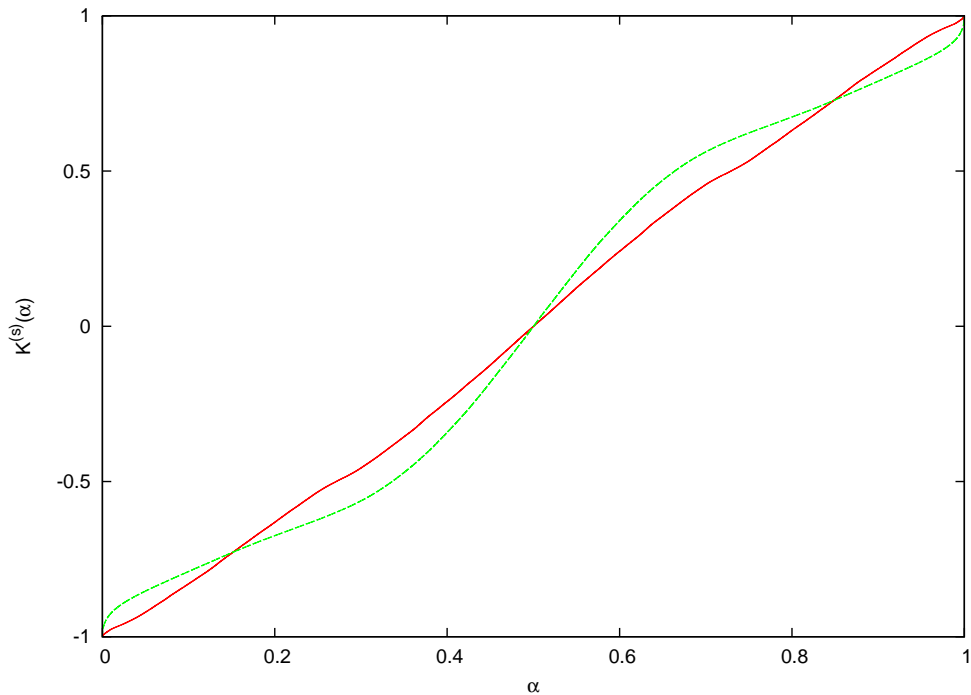}
\caption{
As in Fig.~\ref{figureK128}, but with $N=2048$.
\label{figureK2048} }
\end{figure}

For both samples, and in general for all samples we have analyzed, the
curve varies smoothly between $-1$ and $1$. This implies that $\pi(X)$
is a very broad function, which takes values of order one in its whole
support. That means that large random correlations show up in the mean
field model with a high frequency, their distribution is essentially
uniform, which is a stronger effect than might have been expected.  In
Fig.~\ref{figureK2048} we show the data for the distribution of
correlations in two different samples at $N=2048$ (and, as before,
$T=0.4$). The curves are slightly smoother, but they are not
dramatically different from the case of the smaller, $N=128$
system. As a matter of fact, we believe that $\pi(X)$ is not self
averaging.

A few comments about Fig.~\ref{figureK128} and~\ref{figureK2048} are
in order. As we have already said, the distribution of correlations is
very broad, and its behavior is very different from an exponential
decay: the correlations take all allowed values with very similar
probabilities.  The functional behavior is indeed very close to a
linear one, but with sample to sample fluctuations that are clearly
visible even at $N=2048$.

In this case the shape of the distribution of correlations is
obviously determined by the presence of ``many states'', and we have
very clearly detected their manifestation here.  When we inspect the
whole phase space, the connected correlation functions coincide with
the full correlation functions at all temperature, since, because of
the ${\cal Z}_2$ symmetry of the Hamiltonian, all the magnetizations
$m_{i}^{(s)}$ in sample $s$ are zero within errors  (as mentioned before): we have
verified that this is very well realized in our statistical sample.

A remarkable feature of Fig.~\ref{figureK128} and~\ref{figureK2048} is
the precise symmetry about the value $\Kalpha = 0$. 
More remarkably, the deviations from the symmetry are much smaller than
the disorder sample to disorder sample fluctuations.
This is true with
a very high degree of accuracy in all the disorder samples we have
inspected. That such a symmetry must be present after averaging over
the disorder is obvious, because of the gauge symmetry, involving both
couplings and spins, of spin glasses~\cite{TOULOU1977,MAPARI}. It is,
however, unclear how symmetric a finite size sample should be. In the
large $N$ limit the symmetry should hold for a typical
system~\cite{Derrida}.

We have checked this symmetry property on small systems averaging over
all spin configurations: we have enumerated all coupling
configurations for $N=5$ and we have analyzed a number of coupling
configurations for $N$ up to $21$. The symmetry clearly does not hold
for the (quite atypical) ferromagnetic configuration, where all
$J_{i,j}$ are equal to one.  The symmetry is not
exact even for balanced configurations (with the
same number of positive and negative couplings), but for increasing $N$ 
these balanced configurations turn out
indeed to be more and more symmetric already for small $N$ values,
confirming our findings: the fact that a typical disorder
configuration is, for $N\to\infty$, symmetric, manifests itself with
great accuracy already for small $N$.

\subsection{The correlation functions in one valley\label{SS:ONE}}

We have discussed in the former paragraphs how correlation functions
behave when thermally averaged over all phase space, i.e. allowing the
system to visit all possible states. To better understand the slow
decay (or, in other terms, the wide support of the probability
density that we have observed), it is interesting to determine
what happens when we constrain the dynamics to visit a single, given
state only. As we have already explained, we do that (empirically) by
running a local, slow dynamics, starting from spin configurations at
thermal equilibrium.  

\begin{figure}
\centering \includegraphics[width=0.6\textwidth,angle=270]{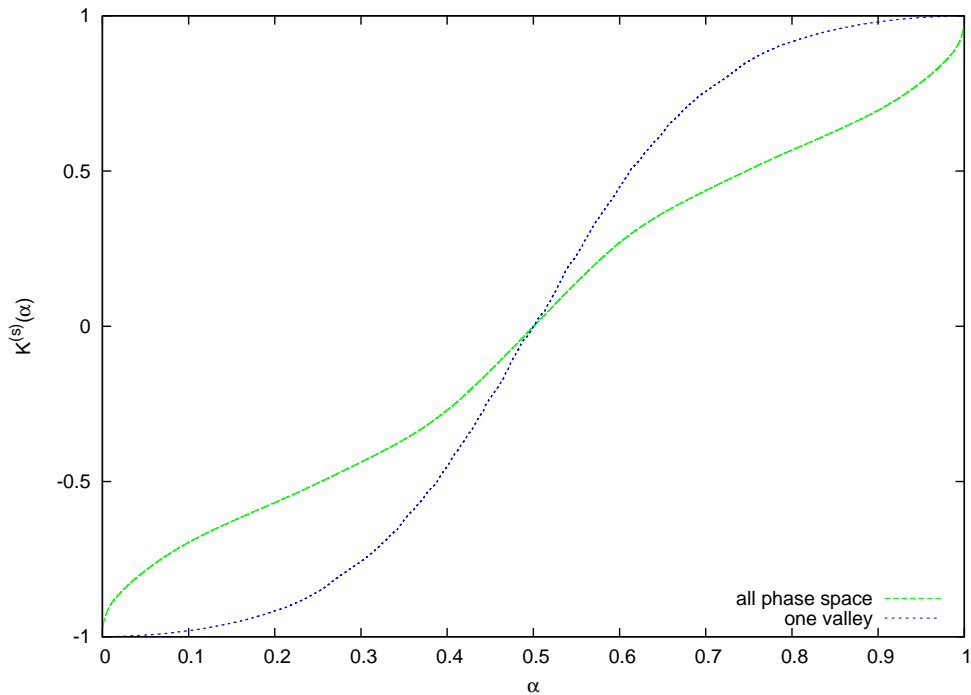}
\caption{(color on line) ${\cal K}^{(s)}(\alpha)$ as a function of
  $\alpha$, for $N=2048$ and $T=0.4$.  The curve that is higher in the 
  left half of the plot is for
  $\Kalpha$ computed by averaging over all states (like in
  Fig.~\ref{figureK2048}), while the curve that is higher in the right part 
  of the plot is for correlation
  functions computed averaging over a single valley.
  \label{figureK2048onevalley} }
\end{figure}

We show in Fig.~\ref{figureK2048onevalley} $\Kalpha$ as a function of
$\alpha$, for a system with $N=2048$ and low $T=0.4$, both as the
result of integrating over all phase space and when it has been
computed inside a given equilibrium free energy valley.

The situation depicted in Fig.~\ref{figureK2048onevalley} is typical
of all samples with more than one valley (in samples where there is a
single valley, i.e. that look paramagnetic, the two correlation
functions are very similar). The single valley correlation functions
are, in this typical situation, of the same shape as the complete
ones, but the difference is obvious (a low value of $T$ is needed to
observe it: we can only resolve this effect clearly at
$T=0.4$, the lowest temperature value we analyze).

The one valley correlation, even when different from the complete
correlation functions, have a finite support where their probability
is significantly different from zero. They are always more biased toward
$\pm 1$ than the complete function.

\subsection{The connected correlation functions\label{SS:CON}}

The one valley correlation functions are indeed in at least one sense
very different from the complete correlation functions: here the site
dependent magnetizations $m_i$ are generically non zero (while when
exploring the full phase space they are all zero). In this case the
connected correlation functions ${\cal K}_{(c)\;ij}$ can be very
different from the simple spin-spin function ${\cal K}_{ij}$.  The
question is how different they are?  Since we are now exploring a
single state, we would expect, far from criticality, a very localized
distribution of the connected correlations and, looking at the
correlations ordered by rank, a very fast, exponential decay.

\begin{figure}
\centering \includegraphics[width=0.6\textwidth,angle=270]{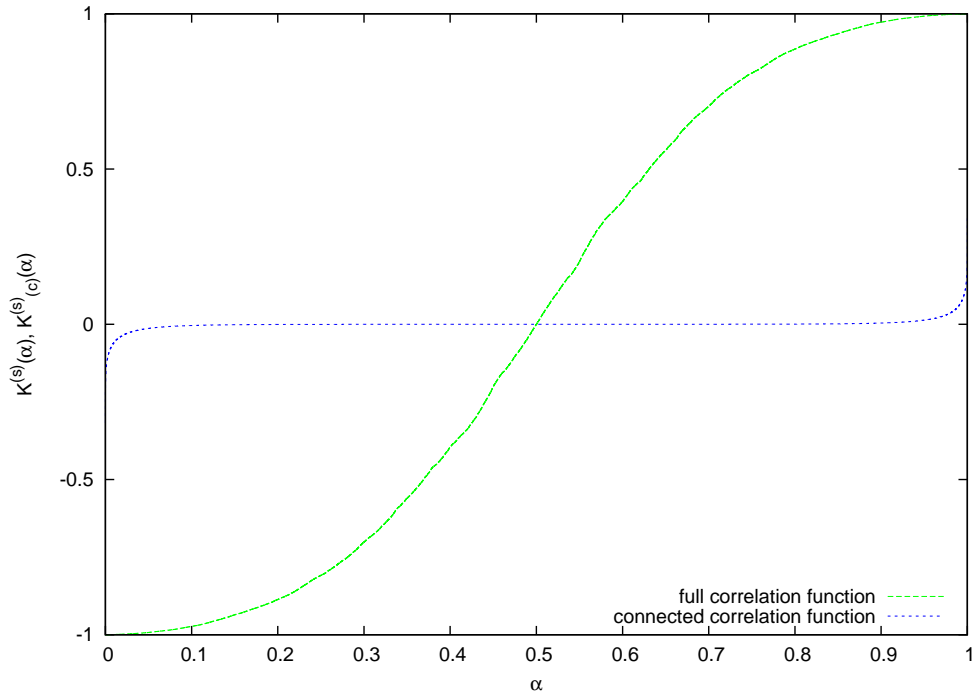}
\caption{The full correlation function ${\cal K}(\alpha)$ and the
  connected correlation function ${\cal K}_{(c)}(\alpha)$ as a
  function of $\alpha$ for a single disorder sample.
\label{figureK2048conn} }
\end{figure}

We have computed these functions for a few samples, $N=2048$ and
different temperature values, down to $T=0.4$. We plot in
Fig.~\ref{figureK2048conn} both the full correlation function ${\cal
  K(\alpha)}$ and the connected correlation 
function ${\cal K}_{(c)}(\alpha)$ as a
function of $\alpha$ for a single disorder sample.

Here the decay is clearly different from the former cases. The
distribution function is indeed very localized, and the plot is surely
indicative of an exponential decay of correlations. 

\section{Conclusions\label{S:CON}}

As we argued in the Introduction, the very concept of complex systems
as being ``more than the sum of their parts" implies that they must
possess long range correlations between their elements. Spin glasses,
with their competing interactions and related complicated phase
space structure, offered themselves as a natural laboratory to test
this idea. In fact, the correlation functions in finite dimensional
spin glasses, averaged over the random couplings, have long been known
to be of long range, both from analytic and from numerical works. Here
we argued that the individual samples must also display interesting
long range correlation structures. As a very first step, we targeted
the mean field model of spin glasses. This choice is motivated by the
fact that the SK model is well understood, and while we are looking
for an unusual, little investigated phenomenon, we are at least moving
on familiar ground. Defined on the complete graph, the SK model does
not have a concept of distance, which is an obvious drawback if we are
to study long range correlations. However, as the correlations in the
individual samples behave in a random, chaotic way even in finite
dimensional spin glasses, the usual representation of correlation
functions as functions of the distance is not very useful anyhow, and
a good global characterization can be obtained by sorting all the
$N(N-1)/2$ correlation coefficients according to magnitude, that is
constructing the probability distribution of correlations. This
construction can be taken over to the mean field model, and this was
the object we have studied here under various conditions. The long
range character of correlations was expected to manifest itself in the
broadening of the distribution of correlations as we go into the
ordered phase. We found that the effect is strikingly strong: when sampling 
the whole phase space the distribution of correlations 
turned out to be essentially uniform at low temperatures, that is any
value between -1 and +1 appeared with basically the same probability,
modulo small scale sample to sample fluctuations. While the other
features we studied (size dependence, temperature dependence,
connected versus full correlations, whole phase space versus
individual valleys) worked out as expected on the basis of the
many valleys picture, the extremely broad distribution of correlations
is a surprise and this result stands out as the central message of
this paper. It is an extra bonus that the second moment of this
distribution turns out to be the same as the second moment of the
overlap distribution, so the broadening of the distribution of
correlations also signals the onset of the splitting of phase space
into many pure states. 
Preliminary investigation \cite{unp}
of spin glass correlations in other topologies
(low dimensional Euclidean lattices and other graphs) indicates that a
definite broadening of the probability density of correlations (albeit
weaker than in the mean field case) is present also in these systems,
that lends support to the conjecture that long range correlations are
a general feature of many, perhaps all, complex systems. We intend to
return to the problem of long range spin glass correlations in finite
dimensions in a subsequent publication.

\section*{Acknowledgments}

AB thanks Bernard Derrida and Yves Le Jan for discussions.  IK was
partially supported by the Teller Program of the National Office for
Research and Technology under grant No. KCKHA005.  He also thanks the
Laboratoire Math\'ematiques Appliqu\'ees aux Syst\`emes, \'Ecole
Centrale de Paris, for the hospitality extended to him during the
final stage of this work.




\section*{References}
\begin{thebibliography}{10}

\bibitem{MEPAVI}
  M. M\'ezard, G. Parisi and M. Virasoro,
  \textit{Spin Glass Theory and Beyond}
  (World Scientific, Singapore 1987).

\bibitem{SK}
  D. Sherrington and S. Kirkpatrick,
  Phys. Rev. Lett. \textbf{35}, 1792 (1975).

\bibitem{DDK}
  C. De Dominicis and I. Kondor,
 Phys. Rev. B \textbf{27}, 606 (1983).

\bibitem{DDK84}
  C. De Dominicis and I. Kondor,
 J. de Physique Lett. \textbf{45}, L205 (1984).
 
\bibitem{DDK85}
  C. De Dominicis and I. Kondor,
  J. de Physique Lett. \textbf{46}, L1037 (1985).

\bibitem{KDD86}
I. Kondor and C. De Dominicis,
Europhys. Lett. \textbf{2}, 617 (1986).

\bibitem{DDKOTE}
T. Temesv\'{a}ri, I. Kondor and C. De Dominicis, 
J. Phys. A: Math. Gen. \textbf{21} L1145 (1988).

\bibitem{DDKOTE98}
C. De Dominicis, I. Kondor and T. Temesv\'{a}ri, {\em Beyond the Sherrington-Kirkpatrick model}, in \textit{Spin glasses and random fields}, edited by A.P. Young, World Scientific, Singapore,  119-160, 1998.

\bibitem{Appropriate} 
E. Marinari, G. Parisi, and J. Ruiz-Lorenzo, 
Phys. Rev. B \textbf{58}, 14852 (1998);
C. De Dominicis, I.Giardina, E. Marinari, O.C. Martin, and F. Zuliani,
Phys. Rev. B \textbf{72}, 014443 (2005);
F. Belletti et al. (the Janus Collaboration) 
Phys. Rev. Lett. \textbf{101}, 157201 (2008); 
J. Stat. Phys. \textbf{135}, 1121 (2009);
P. Contucci, C. Giardin\`a, C. Giberti, G. Parisi and C. Vernia, 
Phys. Rev. Lett. \textbf{103}, 017201 (2009);
R. Alvarez-Banos et al. (the Janus collaboration),
J. Stat. Mech. P06026 (2010);
Phys. Rev. Lett. {\bf 105}, 177202 (2010).

\bibitem{FHBM}
D.S. Fisher and D. Huse, Phys. Rev. Lett. \textbf{56}, 1601 (1986);
Phys. Rev. B \textbf{38}, 386 (1988); A.J. Bray and M.A. Moore in \textit{Proceedings of the Heidelberg Colloquium on Glassy Dynamics} (Lecture Notes in Physics 275), edited by J.L. van Hemmen and I. Morgenstern (Springer, Heidelberg, 1986).

\bibitem{SICAMA2000} 
  J. Sinova, G. Canright, and A. H. MacDonald,
  Phys. Rev. Lett. \textbf{85}, 2609 (2000).

\bibitem{SCCM2001} 
  J. Sinova, G. Canright, H. E. Castillo, and A. H. MacDonald, 
  Phys. Rev. B \textbf{63}, 104427 (2001).

\bibitem{HUKIBA2002}
  K. Hukushima and Y. Iba, 
preprint cond-mat/0207123 (July 2002).

\bibitem{COMAMA2002} 
  L. Correale, E. Marinari, and V. Martin-Mayor,
  Phys. Rev. B \textbf{66}, 174406 (2002).

\bibitem{KOZJ} I. Kondor, \textit{Correlations in Complex Systems},
  invited talk at the International Workshop on Challenges and Visions
  in the Social Sciences, ETH Zurich, August 18-23, 2008;
  \textit{Strong Random Correlations in Complex Systems}, invited talk
  at the 4th European PhD Complexity School, The Hebrew University,
  Jerusalem, September 10-14, 2008.

\bibitem{NS}
C.M. Newman and D.L. Stein,
Phys. Rev. B \textbf{46}, 973  (1992);
A. Billoire and E. Marinari,
Europhys. Lett. {\bf 60}, 775 (2002);
J. Lukic, E. Marinari, O. C. Martin and S. Sabatini,
J. Stat. Mech. L10001 (2006).

\bibitem{ARABER09}
N. Aral and A.N. Berker, 
Phys. Rev. B \textbf{79}, 014434 (2009).

\bibitem{TOULOU1977}
  G. Toulouse, 
  Commun. Phys. (London) \textbf{2}, 115 (1977).

\bibitem{MAPARI}
    E.~Marinari, C.~Parrinello and R.~Ricci,
    Nucl. Phys. B {\textbf 362}, 487 (1991).
    
\bibitem{PT} K. Hukushima and K. Nemoto, J. Phys. Soc. Japan {\textbf
  65}, 1604 (1996); M.C. Tesi, E.J. Janse van Rensburg, E. Orlandini
  and S.G.~Whittington, J. Stat. Phys. {\textbf 82}, 155 (1996).

\bibitem{PT1} E. Marinari, {\em Optimized Monte Carlo Methods} in {\em
  Advances in Computer Simulation}, edited by J.Kertesz and I. Kondor,
  Springer-Verlag (1997).

\bibitem{Derrida} B. Derrida, private communication.

\bibitem{Young} A. Billoire and E. Marinari: J. Phys. A
  Math. Gen. {\bf 34}, L727 (2001); A. Billoire, preprint
  arXiv:1003.6086 [cond-mat] (March 2010).

\bibitem{unp} I. Kondor, I. Csabai and E. Mones, unpublished.
 
\end{thebibliography}
\end{document}